\documentclass[11pt]{article}
\usepackage{geometry}                
\usepackage{graphicx}
\usepackage{amssymb, amsmath}
\usepackage{algorithm, algorithmic}
\usepackage{url}
\usepackage{subfigure}
\title{Nested Sampling}
\author{Michael Betancourt}
\date{}

\begin{document}

\begin{centering}

{\large Nested Sampling with Constrained Hamiltonian Monte Carlo}

\hspace{5mm}

Michael Betancourt \footnote{betan@mit.edu}

\textit{Massachusetts Institute of Technology, Cambridge, MA 02139}

\end{centering}

\hspace{3mm}

\begin{abstract}

Nested sampling is a powerful approach to Bayesian inference ultimately limited by the computationally demanding task of sampling from a heavily constrained probability distribution.  An effective algorithm in its own right, Hamiltonian Monte Carlo is readily adapted to efficiently sample from any smooth, constrained distribution.  Utilizing this constrained Hamiltonian Monte Carlo, I introduce a general implementation of the nested sampling algorithm.

\end{abstract}

\section{Bayesian Inference}

Bayesian inferenceis a diverse and robust analysis methodology  \cite{MacKay2003,Jaynes2003}  based on Bayes' Theorem,

\begin{equation*}
p \left( \alpha | \mathcal{D}, H \right) = \frac{ p \left( \mathcal{D} | \alpha, H \right) p \left( \alpha | H \right) }{ p \left( \mathcal{D} | H \right) }.
\end{equation*}

The prior,

\begin{equation*}
p \left( \alpha | H \right) \equiv \pi \left( \alpha \right),
\end{equation*}

\noindent encodes all knowledge about the parameters $\alpha$ before the data $\mathcal{D}$ have been collected, while the likelihood,

\begin{equation*}
p \left( \mathcal{D} | \alpha, H \right) \equiv \mathcal{L} \left( \alpha \right),
\end{equation*}

\noindent defines the probabilistic model of how the data are generated.  The evidence,

\begin{equation*}
p \left( \mathcal{D} | H \right) = \int \mathrm{d}^{m} \alpha \, \mathcal{L} \left( \alpha \right) \pi \left( \alpha \right) \equiv Z,
\end{equation*}

\noindent ensures proper normalization while allowing for auxiliary applications such as model comparison.  Lastly, the posterior,

\begin{equation*}
p \left( \alpha | \mathcal{D}, H \right) \equiv p \left( \alpha \right),
\end{equation*}

\noindent is the refinement of the prior $\pi \left( \alpha \right)$ given the information inferred from $\mathcal{D}$.  All model assumptions are captured by the conditioning hypothesis $H$.

While Bayes' Theorem is simple enough to formulate, in practice the individual components are often sufficiently complex that analytic manipulation is not feasible and one must resort to approximation.  One of the more successful approximation techniques, Markov Chain Monte Carlo (MCMC) produces samples directly from the posterior distribution that are often sufficient to characterize even high dimensional distributions.  The one manifest limitation of MCMC, however, is the inability to directly calculate the evidence $Z$, which, as MacKay notes, ``is often the single most important number in the problem'' \cite{MacKay2003}.

Nested sampling \cite{Skilling2004} is an alternative to sampling from the posterior that instead emphasizes the calculation of the evidence.

\begin{figure}
\centering
\subfigure[]{\includegraphics[width=2.3in]{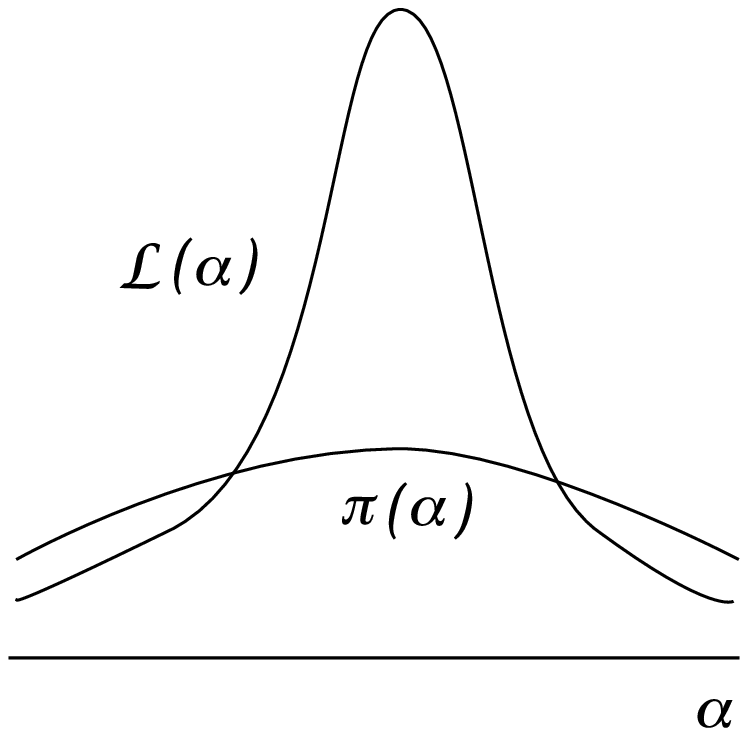}}
\subfigure[]{\includegraphics[width=2.3in]{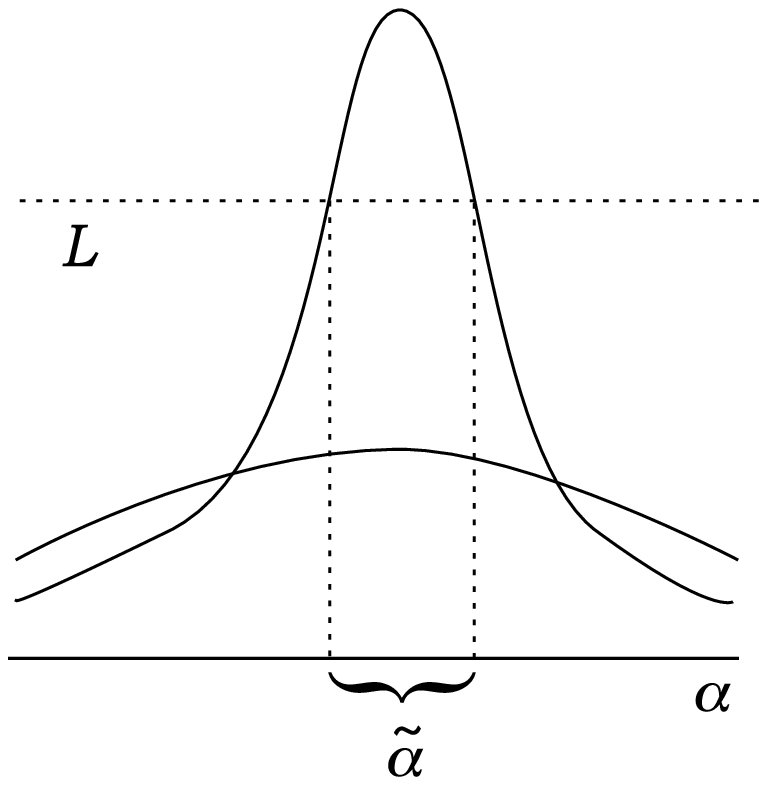}}
\subfigure[]{\includegraphics[width=2.3in]{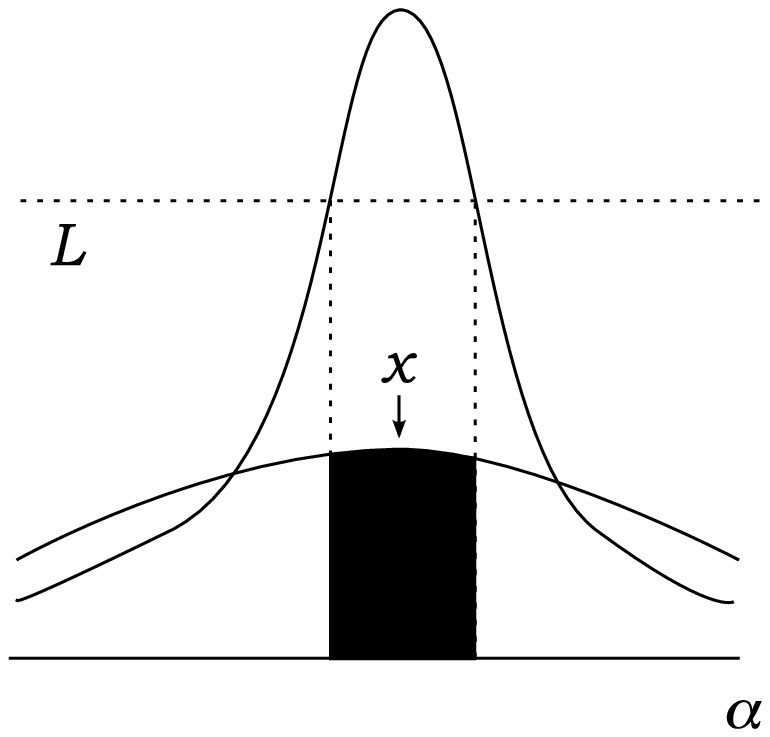}}
\subfigure[]{\includegraphics[width=2.3in]{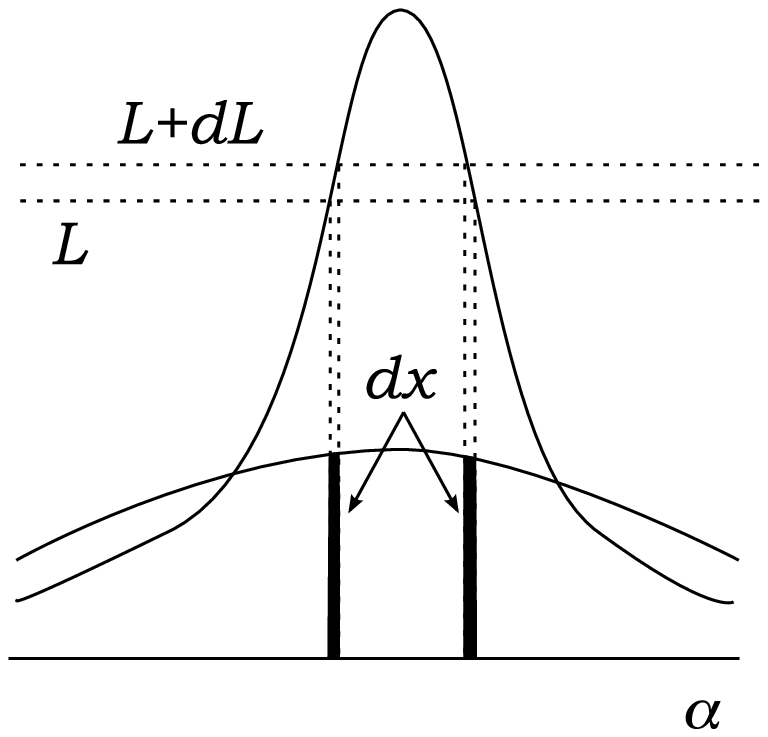}}
\caption{ (a) The prior distribution and likelihood, (b) $\tilde{\alpha}$, the set of $\alpha$ satisfying the likelihood bound $\mathcal{L} \left( \alpha \right) > L$, (c) the prior mass $x\left(L\right)$, and (d) the differential prior mass $\mathrm{d}x$.
\label{fig:nestedSampling}}
\end{figure}

\begin{figure}
\centering
\includegraphics[width=2.5in]{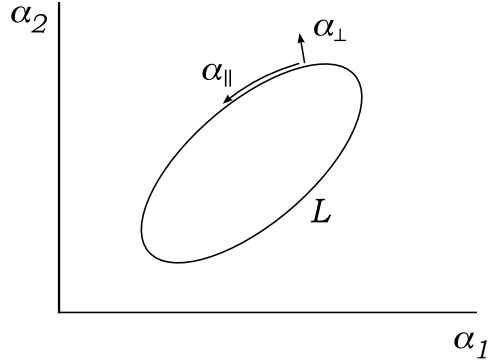}
\caption{ Cartoon of $\alpha_{\perp}$ and $\alpha_{\parallel}$ in two dimensions.   $\alpha_{\parallel}$ parameterizes the contour of constant likelihood $L$ while $\alpha_{\perp}$ parameterizes translations orthogonal to the contour.
\label{fig:alphaRepar}}
\end{figure}

\section{Nested Sampling}

Consider the support of the likelihood above a given bound $L$ (Fig \ref{fig:nestedSampling}a, \ref{fig:nestedSampling}b), 

\begin{equation*}
\tilde{\alpha} = \{ \alpha | \mathcal{L} \left( \alpha \right) > L \},
\end{equation*}

\noindent and the associated prior mass across that support (Fig \ref{fig:nestedSampling}c),

\begin{equation*}
x \left( L \right) = \int_{\tilde{\alpha}} \mathrm{d}^{m} \alpha \, \pi \left( \alpha \right).
\end{equation*}

The differential $\mathrm{d}x$ gives the prior mass associated with the likelihood $L = \mathcal{L} \left( \alpha \right)$ (Fig \ref{fig:nestedSampling}d),

\begin{align*}
\mathrm{d}x \left(L\right) &= \mathrm{d} \int_{\tilde{\alpha}} \mathrm{d}^{m} \alpha \, \pi \left( \alpha \right) \\
\mathrm{d}x \left(L\right) &= \int_{\partial \tilde{\alpha}} \mathrm{d}^{m} \alpha \, \pi \left( \alpha \right)
\end{align*}

\noindent where $\partial \tilde{\alpha}$ is the $m-1$ dimensional boundary of constant likelihood,

\begin{equation*}
\partial \tilde{\alpha} = \{ \alpha | \mathcal{L} \left( \alpha \right) = L \}.
\end{equation*}

Introducing the coordinate $\alpha_{\perp}$ perpendicular to the likelihood constraint boundary and the $m-1$ coordinates $\alpha_{\parallel}$ parallel to the constraint (Fig \ref{fig:alphaRepar}), the integral over $\partial \tilde{\alpha}$ simply marginalizes $\alpha_{\parallel}$ and the differential becomes

\begin{align*}
\mathrm{d}x \left(L\right) &= \int_{\partial \tilde{\alpha}} \mathrm{d} \alpha_{\perp} \mathrm{d}^{m -1} \alpha_{\parallel} \, \pi \left( \alpha \right) \\
\mathrm{d}x \left(L\right) &= \mathrm{d} \alpha_{\perp}  \int_{\partial \tilde{\alpha}} \mathrm{d}^{m -1} \alpha_{\parallel} \, \pi \left( \alpha \right) \\
\mathrm{d}x \left(L\right) &= \mathrm{d} \alpha_{\perp}  \pi \left( \alpha_{\perp} \right).
\end{align*}

\pagebreak

Returning to the evidence,

\begin{align*}
Z &=  \int \mathrm{d}^{m} \alpha \, \mathcal{L} \left( \alpha \right) \pi \left( \alpha \right) \\
Z &=  \int \mathrm{d} \alpha_{\perp} \mathrm{d}^{m-1} \alpha_{\parallel} \, \mathcal{L} \left( \alpha \right) \pi \left( \alpha \right).
\end{align*}

\noindent By construction the likelihood is invariant to changes in $\alpha_{\parallel}$, $\mathcal{L} \left( \alpha \right) = \mathcal{L} \left( \alpha_{\perp} \right)$, and the integral simplifies to

\begin{align*}
Z &=  \int \mathrm{d} \alpha_{\perp} \mathrm{d}^{m-1} \alpha_{\parallel} \, \mathcal{L} \left( \alpha_{\perp} \right) \pi \left( \alpha \right) \\
Z &=  \int \mathrm{d} \alpha_{\perp} \mathcal{L} \left( \alpha_{\perp} \right) \int \mathrm{d}^{m-1} \alpha_{\parallel} \,  \pi \left( \alpha \right) \\
Z &=  \int \mathrm{d} \alpha_{\perp} \mathcal{L} \left( \alpha_{\perp} \right) \pi \left( \alpha_{\perp} \right) \\
Z &=  \int \mathrm{d} \alpha_{\perp} \pi \left( \alpha_{\perp} \right) \mathcal{L} \left( \alpha_{\perp} \right) \\
Z &=  \int \mathrm{d} x \, L \left( x \right)
\end{align*}

\noindent where $L \left( x\right) = \mathcal{L} \left( \alpha_{\perp} \left(x\right) \right)$ is the likelihood bound resulting in the prior mass $x$.

This clever change of variables has reduced the $m$ dimensional integration over the parameters $\alpha$ to a one dimensional integral over the bounded support of $x$.   Although this simplified integral is easier to calculate in theory, it is fundamentally limited by the need to compute $L\left(x\right)$.  

Numerical integration, however, needs only a set of points $\left( x_{k}, L_{k} \right)$ and not $L \left(x\right)$ explicitly.  Sidestepping $L \left(x\right)$, consider instead the problem of generating the set $\left( x_{k}, L_{k} \right)$ directly.

\subsection{Sampling $L\left(x\right)$ With The Shrinkage Distribution}

In particular, consider a stochastic approach beginning with $n$ samples drawn from $\pi \left( \alpha \right)$.  The sample with the smallest likelihood, $\mathcal{L}_{\min}$, bounds the largest $x$ but otherwise nothing can be said of the exact value, $x_{\max}$, without an explicit, and painful, calculation from the original definition. 

The cumulative probability of $x_{\max}$, however, is simply the probability of $x_{\max}$ exceeding the $x$ of each sample,

\begin{align*}
P \left( x_{\max} \right) &= P \left( x_{1} \le x_{\max} \right) \cdots P \left( x_{n} \le x_{\max} \right) \\
P \left( x_{\max} \right) &= \int_{0}^{x_{\max}} \mathrm{d}x \, \pi \left(x\right)  \cdots \int_{0}^{x_{\max}} \mathrm{d}x \, \pi \left(x\right) \\
P \left( x_{\max} \right) &= \left( \int_{0}^{x_{\max}} \mathrm{d}x \, \pi \left(x\right) \right)^{n},
\end{align*}

\noindent where $\pi \left(x \right)$ is uniformly distributed,

\begin{align*}
\pi \left(x \right) &= \int_{\partial \tilde{\alpha}} \mathrm{d}^{m-1} \alpha_{\parallel} \, \pi \left( \alpha \left(x \right) \right) \left| \frac{\mathrm{d}\alpha}{\mathrm{d}x} \right| \\
\pi \left(x \right) &= \int_{\partial \tilde{\alpha}} \mathrm{d}^{m-1} \alpha_{\parallel} \, \pi \left( \alpha \left(x \right) \right) \left| \frac{\mathrm{d}\alpha_{\perp}}{\mathrm{d}x} \right| \\
\pi \left(x \right) &= \int_{\partial \tilde{\alpha}} \mathrm{d}^{m-1} \alpha_{\parallel} \, \pi \left( \alpha \left(x \right) \right) \frac{1}{ \pi \left( \alpha_{\perp} \left(x\right) \right) } \\
\pi \left(x \right) &=  \frac{1}{ \pi \left( \alpha_{\perp} \left(x\right) \right) } \int_{\partial \tilde{\alpha}} \mathrm{d}^{m-1} \alpha_{\parallel} \, \pi \left( \alpha \left(x \right) \right) \\
\pi \left(x \right) &=  \frac{1}{ \pi \left( \alpha_{\perp} \left(x\right) \right) } \pi \left( \alpha_{\perp} \left(x \right) \right) \\
\pi \left(x \right) &= \left\{ \begin{array}{rc} 1, & 0 \leq x \leq 1 \\ 0, & \mathrm{otherwise} \end{array} \right. 
\end{align*}

\pagebreak

\noindent Simplifying, the cumulative probability of the largest sample reduces to

\begin{align*}
P \left( x_{\max} \right) &= \left( \int_{0}^{x_{\max}} \mathrm{d}x \, \pi \left(x\right) \right)^{n} \\
P \left( x_{\max} \right) &= \left( \int_{0}^{x_{\max}} \mathrm{d}x \right)^{n} \\
P \left( x_{\max} \right) &= x_{\max}^{n}
\end{align*}

\noindent with the corresponding probability distribution

\begin{align*}
p \left( x_{\max} \right) &= \frac{\mathrm{d} P \left( x_{\max} \right) }{\mathrm{d}x_{\max}} \\
p \left( x_{\max} \right) &= n x_{\max}^{n -1}.
\end{align*}

\noindent Estimating $x_{\max}$ from the probability distribution $p \left( x_{\max} \right)$ immediately yields a pair 

\begin{equation*}
\left( x_{1} = x_{\max}, L_{1} = \mathcal{L}_{\min} \right).
\end{equation*}

A second pair follows by drawing from the constrained prior

\begin{equation*}
\tilde{\pi} \left(\alpha \right) \propto \left\{ \begin{array}{rc} \pi \left(\alpha\right), & \mathcal{L} \left( \alpha \right) > \mathcal{L}_{1} \\ 0, & \mathrm{otherwise} \end{array} \right. ,
\end{equation*}

\noindent or in terms of $x$,

\begin{equation*}
\tilde{\pi} \left(x \right) = \left\{ \begin{array}{rc} 1 / x_{1}, & 0 \leq x \leq x_{1} \\ 0, & \mathrm{otherwise} \end{array} \right. .
\end{equation*}

\noindent $n$ samples from this constrained prior yield a new minimum $L_{2}$ with $x_{2}$ distributed as

\begin{equation*}
p \left( x_{2} | x_{1} \right) = \frac{n}{x_{1}} \left( \frac{x_{2}}{x_{1}} \right)^{n -1}
\end{equation*}

\noindent Making another point estimate gives $\left(x_{2}, L_{2} \right)$.

Generalizing, the $n$ samples at each iteration are drawn from a uniform prior restricted by the previous iteration,

\begin{equation*}
\tilde{\pi} \left(x \right) = \left\{ \begin{array}{rc} 1 / x_{k-1}, & 0 \leq x \leq x_{k-1} \\ 0, & \mathrm{otherwise} \end{array} \right. ,
\end{equation*}

\noindent The distribution of the largest sample, $x_{k}$, follows as before,

\begin{equation*}
p \left( x_{k} | x_{k-1} \right) = \frac{n}{x_{k-1}} \left( \frac{x_{k}}{x_{k-1}} \right)^{n -1},
\end{equation*}

Note that this implies that the shrinkage at each iteration, $t_{k} = x_{k} / x_{k - 1}$, is identically and independently distributed as

\begin{equation*}
p \left( t_{k} \right) = p \left(t\right) = n t_{k}^{n-1}.
\end{equation*}

\noindent Moreover, a point estimate for $x_{k}$ can be written entirely in terms of point estimates for the $t_{k}$,

\begin{align*}
x_{k} &= \frac{x_{k}}{x_{k-1}} \cdot \frac{x_{k-1}}{x_{k-2}} \ldots \frac{x_{2}}{x_{1}} \cdot \frac{x_{1}}{x_{0}} \cdot x_{0} \\
x_{k} &= t_{k} \cdot t_{k - 1} \ldots t_{2} \cdot t_{1} \cdot x_{0} \\
x_{k} &= \left( \prod_{i = 1}^{k} t_{i} \right) x_{0}.
\end{align*}

More appropriate to the large dynamic ranges encountered in many applications, $\log x_{k}$ becomes

\begin{align*}
\log x_{k} &= \log \left( \prod_{i = 1}^{k} t_{i} \right) x_{0} \\
\log x_{k} &= \sum_{i = 1}^{k} \log t_{i} + \log x_{0}.
\end{align*}

Performing a quick change of variables, the logarithmic shrinkage will be distributed as

\begin{equation*}
p \left( \log t \right) = n e^{n \log t}
\end{equation*}

\noindent with the mean and standard deviation

\begin{equation*}
\log t = - \frac{1}{n} \pm \frac{1}{n}.
\end{equation*}

Taking the mean as the point estimate for each $\log t_{i}$ finally gives

\begin{equation*}
\log x_{k} - \log x_{0} = -\frac{k}{n}
\end{equation*}

\noindent with the resulting error

\begin{equation*}
\delta \left( \log x_{k} - \log x_{0} \right) = \frac{\sqrt{k}}{n}.
\end{equation*}

\noindent Parameterizing $x_{k}$ in terms of the shrinkage proves immediately advantageous -- because the $\log t_{i}$ are independent, the errors in the point estimates tend to cancel and the estimate for the $x_{k}$ grow increasingly more accurate with $k$.

At each iteration, then, a pair $\left( x_{k}, L_{k} \right)$ is given by the point estimate for $x_{k}$ and the smallest likelihood of the $n$ drawn samples.

\subsection{The Algorithm}

\begin{figure}
\centering
\subfigure[]{\includegraphics[width=2.5in]{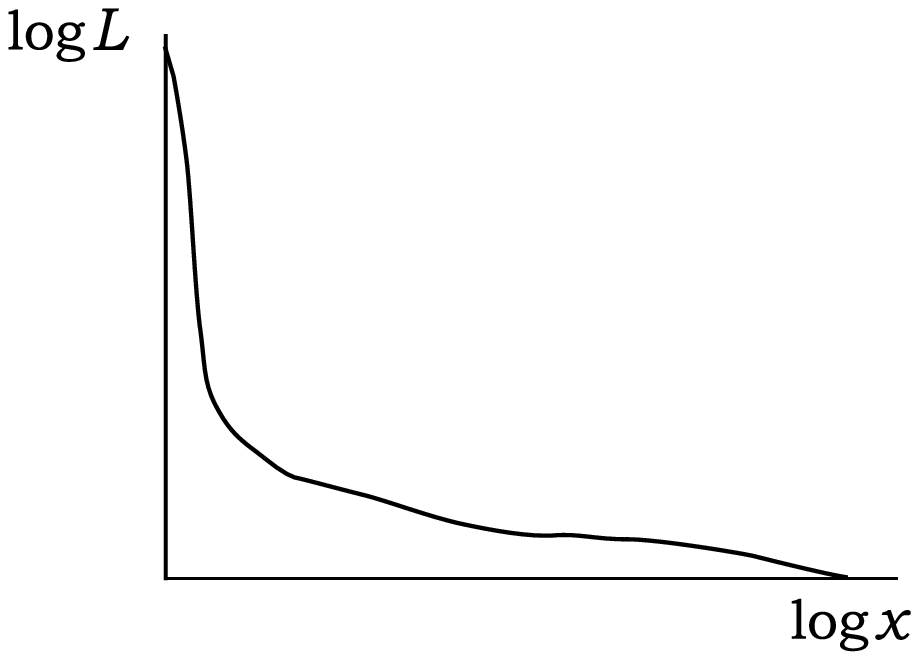}}
\subfigure[]{\includegraphics[width=2.5in]{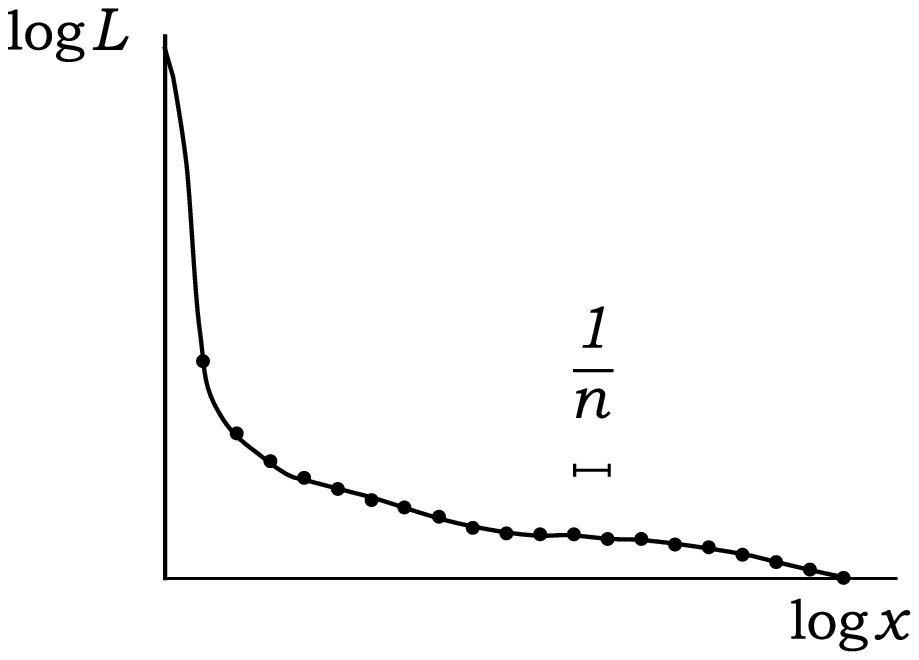}}
\caption{ (a) The evidence integrand $L\left(x\right)$.  Note how the bulk resides at exponentially small values of $x$. (b) Taking the mean of the shrinkage distribution, nested sampling generates a series of samples $\left(x_{k}, L_{k} \right)$ with $\log x_{k - 1} - \log x_{k} = 1 / n$.
\label{fig:nestedSamples}}
\end{figure}

A proper implementation of nested sampling begins with the initial point $\left( x_{0} = 1, L_{0} = 0 \right)$.  At each iteration, $n$ samples are drawn from the constrained prior

\begin{equation*}
\tilde{\pi} \left(\alpha \right) \propto \left\{ \begin{array}{rc} \pi \left(\alpha\right), & \mathcal{L} \left( \alpha \right) > \mathcal{L}_{k-1} \\ 0, & \mathrm{otherwise} \end{array} \right. 
\end{equation*}

\noindent and the sample with the smallest likelihood provides a ``nested'' sample with $L_{k} = \mathcal{L} \left( \alpha_{k} \right)$ and $\log x_{k} = -\frac{k}{n}$ (Figure \ref{fig:nestedSamples}).  $\mathcal{L} \left( \alpha_{k} \right)$ defines a new constrained prior for the following iteration.  Note that the remaining samples from the given iteration will already satisfy this new likelihood constraint and qualify as $n - 1$ of the samples necessary for the next iteration -- only one new sample will actually need to be generated.

As the algorithm iterates, regions of higher likelihood are reached until the nested samples begin to converge to the maximum likelihood.  Determining this convergence is tricky, but heuristics have been developed that are quite successful for well behaved likelihoods \cite{Skilling2004,Sivia2006}.

Once the iterations have terminated, the evidence is numerically integrated using the nested samples. The simplest approach is a first order numerical quadrature:

\begin{align*}
Z &\approx \sum_{k} \left( x_{k-1} - x_{k} \right) L_{k} \\
Z &\approx \sum_{k} \left( e^{\log x_{k-1}} - e^{\log x_{k}} \right) L_{k} \\
Z &\approx \sum_{k} \left( e^{- \frac{k-1}{n} } - e^{ - \frac{k}{n}} \right) L_{k} \\
Z &\approx \sum_{k} e^{- \frac{k - 1}{n}} \left( 1 - e^{ - \frac{1}{n}} \right) L_{k}.
\end{align*}

\noindent Errors from the numerical integration are dominated by the errors from the use of point estimates and, consequently, higher order quadrature offers little improvement beyond the first order approximation.

The errors inherent in the point estimates can be reduced by instead marginalizing over the shrinkage distributions.  Note, however, that in many applications the likelihood will be relatively peaked and most of the prior mass will lie within its tails.  $x\left(L\right)$ will then be heavily weighted towards exponentially small values of $L$ where the likelihood constraint falls below the tails and the prior mass rapidly accumulates.  Likewise, the integrand $L\left(x\right)$ will be heavily weighted towards exponentially small values of $x$ and the dominant contributions from the quadrature will come from later iterations, exactly where the point estimates become more precise.  The resulting error in the integration tends to be reasonable, and the added complexity of marginalization offers little improvement.

The choice of $n$ can also be helpful in improving the accuracy of the integration.  For larger $n$ the shrinkage distribution narrows and the estimates for the $x_{k}$ become increasingly better.  Multiple samples at each iteration also prove valuable when the likelihood is multimodal, as the individual samples allow the modes to be sampled simultaneously \cite{Sivia2006}. 

Lastly, if the $\alpha$ yielding the smallest likelihood are stored with each nested sample then posterior expectations can be estimated with the quadrature weights,

\begin{equation*}
\bar{f} = \int \mathrm{d}^{m} \alpha \, p \left( \alpha \right) f \left( \alpha \right) \approx \sum_{k} \frac{ L_{k} \left( x_{k- 1} - x_{k} \right) }{Z} f \left( \alpha_{k} \right).
\end{equation*}

The remaining obstacle to a fully realized algorithm is the matter of sampling from the prior given the likelihood constraint $\mathcal{L} > \mathcal{L}_{\mathrm{min}}$.  Sampling from constrained distributions is a notoriously difficult problem, and recent applications of nested sampling have focused on modifying the algorithm in order to make the constrained sampling feasible \cite{Feroz2008,Brewer2009}.  Hamiltonian Monte Carlo, however, offers samples directly from the constrained prior and provides an immediate implementation of nested sampling.

\section{Hamiltonian Monte Carlo}

Hamiltonian Monte Carlo \cite{MacKay2003,Bishop2007,Neal2010} is an efficient method for generating samples from the $m$ dimensional probability distribution

\begin{equation*}
p \left( \mathbf{x} \right) \propto \exp \left[ - E \left( \mathbf{x} \right) \right].
\end{equation*}

First, consider instead the larger distribution 

\begin{equation*}
p \left( \mathbf{x}, \mathbf{p} \right) = p \left( \mathbf{x} \right) p \left( \mathbf{p} \right)
\end{equation*}

\noindent where the latent variables $\mathbf{p}$ are i.i.d. standardized Gaussians

\begin{equation*}
p \left( \mathbf{p} \right) \propto \exp \left( - \frac{1}{2} \left| \mathbf{p} \right|^{2} \right).
\end{equation*}

The joint distribution of the initial $\mathbf{x}$ and the latent $\mathbf{p}$ is then

\begin{align*}
p \left( \mathbf{x}, \mathbf{p} \right) &\propto \exp \left( - \frac{1}{2} \left| \mathbf{p} \right|^{2} - E \left( \mathbf{x} \right) \right) \\
p \left( \mathbf{x}, \mathbf{p} \right) &\propto \exp \left( - H \right)
\end{align*}

\noindent where

\begin{equation*}
H \equiv \frac{1}{2} \left| \mathbf{p} \right|^{2} + E \left( \mathbf{x} \right)
\end{equation*}

\noindent takes the form of the Hamiltonian of classical mechanics.

Applying Hamilton's equations

\begin{equation*}
\frac{d \mathbf{x} }{dt} = \frac{ \partial H}{\partial \mathbf{p}} = \mathbf{p}
\end{equation*}

\begin{equation*}
\frac{d \mathbf{p} }{dt} = - \frac{ \partial H}{\partial \mathbf{x}} = - \nabla E \left( \mathbf{x} \right)
\end{equation*}

\noindent to a given sample $\{ \mathbf{x}, \mathbf{p} \}$ produces a new sample $\{ \mathbf{x}', \mathbf{p}' \}$.  Note that the properties of Hamiltonian dynamics, in particular Liouville's Theorem and conservation of $H$, guarantee that differential probability masses from $p \left( \mathbf{x}, \mathbf{p} \right)$ are conserved by the mapping.  As a result, this dynamic evolution serves as a transition matrix $T \left(  \mathbf{x}, \mathbf{p} ;  \mathbf{x}', \mathbf{p}' \right)$ with the invariant distribution $p \left( \mathbf{x}, \mathbf{p} \right)$.  Moreover, the time reversal symmetry of the equations ensures that the evolution satisfies detailed balance:

\begin{equation*}
T \left(  \mathbf{x}, \mathbf{p} ;  \mathbf{x}', \mathbf{p}' \right) = T \left(  \mathbf{x}', \mathbf{p}' ;  \mathbf{x}, \mathbf{p} \right).
\end{equation*}

Because $H$ is conserved, however, the transitions are not ergodic and the samples do not span the full support of $p \left( \mathbf{x}, \mathbf{p} \right)$.  Ergodicity is introduced by adding a Gibbs sampling step for the $\mathbf{p}$.  Because the $\mathbf{x}$ and $\mathbf{p}$ are independent, sampling from the conditional distribution for $\mathbf{p}$ is particularly easy 

\begin{align*}
p \left( \mathbf{p} | \mathbf{x} \right) &= p \left( \mathbf{p} \right) \\
p \left( \mathbf{p} | \mathbf{x} \right) &= \prod_{i = 1}^{m} \mathcal{N} \left(0, 1\right).
\end{align*}

The algorithm proceeds by alternating between dynamical evolution and Gibbs sampling and the resulting samples $ \{ \mathbf{x}_{k}, \mathbf{p}_{k} \}$ form a proper Markov chain.

In practice the necessary integration of Hamilton's equations cannot be performed analytically and one must resort to numerical approximations.  Unfortunately, any discrete approximation will lack the symmetry necessary for both Liouville's Theorem and energy conservation to hold, and the exact invariant distribution will no longer be $p \left( \mathbf{x}, \mathbf{p} \right)$.  This can be overcome by treating the evolved sample as a Metropolis proposal, accepting proposed samples with probability

\begin{align*}
P \left( \mathrm{accept} \right) &= \min \left(1, \frac{ p \left( \mathbf{x}', \mathbf{p}' \right) }{ p \left( \mathbf{x}, \mathbf{p} \right) } \right) \\
P \left( \mathrm{accept} \right) &= \min \left(1, \frac{ \exp \left( - H' \right) }{ \exp \left( - H \right) } \right) \\
P \left( \mathrm{accept} \right) &= \min \left(1, \exp \left( - \Delta H \right) \right).
\end{align*}

\noindent All transitions to a smaller Hamiltonian, and hence higher probability, are automatically accepted.  Transitions resulting in a larger Hamiltonian are only occasionally accepted.

Further implementation details, particularly insight on the choice of step size and total number of steps, can be found in \cite{Neal2010}.

\subsection{Constrained Hamiltonian Monte Carlo}

Now consider the constrained distribution

\begin{equation*}
\tilde{p} \left( \mathbf{x} \right) \propto \left\{ 
\begin{array}{rc}
p \left( \mathbf{x} \right),&  C \left( \mathbf{x} \right) \geq 0 \\ 
0, & \mathrm{else}
\end{array} 
\right. .
\end{equation*}

Sampling from $\tilde{p} \left( \mathbf{x} \right)$ is challenging.  The simplest approach is to sample from $p \left( \mathbf{x} \right)$ and discard those not satisfying the constraint.  For most nontrivial constraints, however, this approach is extremely inefficient as the majority of the computational effort is spent generating samples that will be immediately discarded.

\begin{figure}[h]
\centering
\subfigure[]{\includegraphics[width=1.5in]{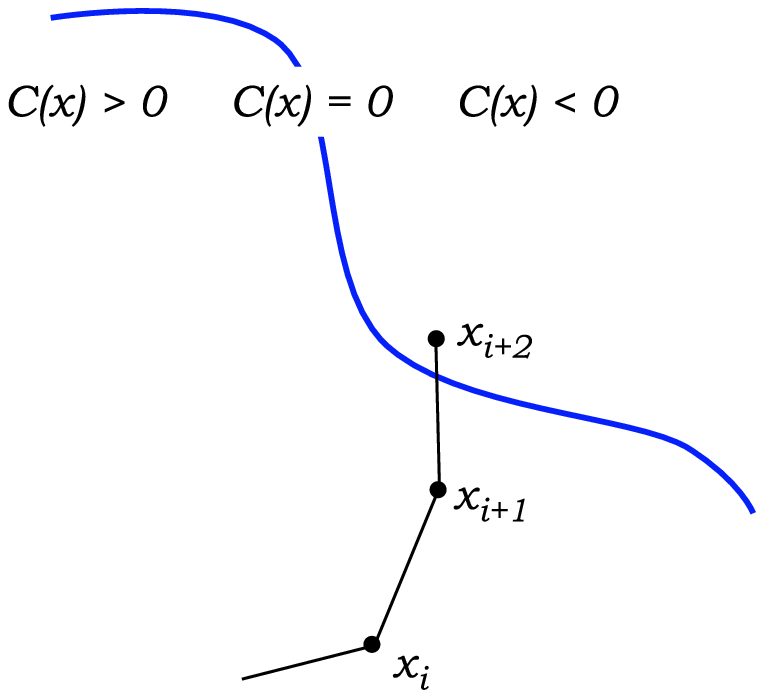}}
\subfigure[]{\includegraphics[width=1.5in]{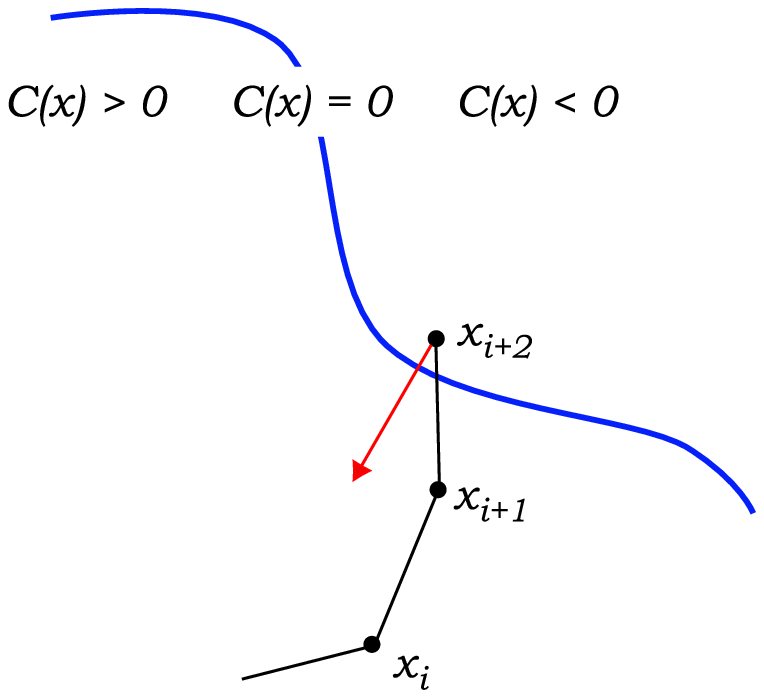}}
\subfigure[]{\includegraphics[width=1.5in]{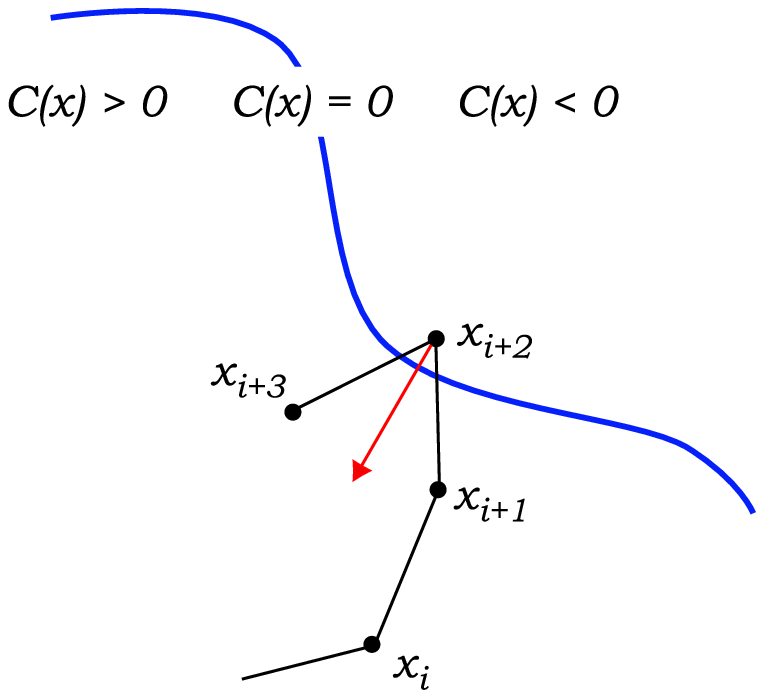}}
\caption{ Cartoon of a particle bouncing off the constraint boundary $C(x) = 0$.  (a) At step $i+2$ the particle violates the constraint, at which point (b) the normal at $x_{i + 2}$ is computed and the momenta reflected in lieu of the normal leapfrog update. (c) The next spatial update is no longer in violation of the constraint.
\label{fig:bounce}}
\end{figure}

From the Hamiltonian point of view, the constraint becomes an infinite potential barrier

\begin{equation*}
\tilde{E} \left( \mathbf{x} \right) = \left\{ 
\begin{array}{rc}
E \left( \mathbf{x} \right), & C \left( \mathbf{x} \right) \geq 0 \\ 
\infty, &\mathrm{else}
\end{array} 
\right. .
\end{equation*}

Incorporating infinite barriers directly into Hamilton's equations is problematic, but physical intuition provides an alternative approach.  Particles incident on an infinite barrier bounce, the momenta perpendicular to the barrier perfectly reflecting:

\begin{align*}
\mathbf{p}' &= \mathbf{p}_{T} - \mathbf{p}_{N} \\
\mathbf{p}' &= \mathbf{p} - 2 \left( \mathbf{p} \cdot \hat{\mathbf{n}} \right) \hat{ \mathbf{n} }.
\end{align*}

\noindent Instead of dealing with infinite gradients, then, one can replace the momenta updates with reflections when the equations integrate beyond the support of $\tilde{p} \left( \mathbf{x} \right)$.

Discrete updates proceed as follows.  After each spatial update the constraint is checked and if violated then the normal $\hat{ \mathbf{n} }$ is computed at the new point and the ensuing momentum update is replaced by reflection (Algo \ref{algo:bounce}, Fig \ref{fig:bounce}).  Note that the spatial update cannot be reversed, nor can an interpolation to the constraint boundary be made, without spoiling the time-reversal symmetry of the evolution.

For smooth constraints $C \left( \mathbf{x} \right) \geq 0$ the normal is given immediately by

\begin{equation*}
\hat{ \mathbf{n} } = \frac{ \nabla C \left( \mathbf{x} \right) } { \left| \nabla C \left( \mathbf{x} \right) \right| }.
\end{equation*}

\noindent The normal for many discontinuous constraints, which are particularly useful for sampling distributions with limited support without resorting to computationally expensive exponential reparameterizations, can be determined by the geometry of the problem.

Finally, if the evolution ends in the middle of a bounce, with the proposed sample laying just outside of the support of $\tilde{p} \left( \mathbf{x} \right)$, it is immediately rejected as the acceptance probability is zero,

\begin{equation*}
P\left(\mathrm{accept}\right) = \exp \left( - \Delta H \right) = \exp \left( - \infty \right) = 0.
\end{equation*}

Given a seed satisfying the constraint, the resultant Markov chain bounces around $\tilde{p} \left( \mathbf{x} \right)$ and avoids the inadmissible regions almost entirely.  Computational resources are spent on the generation of relevant samples and the sampling proceeds efficiently no matter the scale of the constraint.

\begin{algorithm}
\caption{Dynamic evolution with a first order leapfrog discretization of Hamilton's equations and constraint $C(\mathbf{x}) \geq 0$.}
\label{algo:bounce}
\begin{algorithmic}

\STATE \COMMENT{First momentum half step}
\STATE $\mathbf{p} \gets \mathbf{p} - \frac{1}{2} \epsilon \nabla E \left( \mathbf{x} \right)$
\STATE
\FOR{$t = 0$ to $T$}

	\STATE
	\STATE \COMMENT{Full spatial step}
	\STATE $ \mathbf{x} \gets \mathbf{x} + \epsilon \mathbf{p}$
	\STATE
	\STATE \COMMENT{Check for constraint}
		\IF{ $C(\mathbf{x}) \geq 0$ }
		         \STATE \COMMENT{Full momentum step}
			\STATE $ \mathbf{p} \gets \mathbf{p} - \epsilon \nabla E \left( \mathbf{x} \right)$
		\ELSE
		         \STATE \COMMENT{Bounce}
			\STATE $\hat{ \mathbf{n} } \gets \nabla C(\mathbf{x}) / \left| \nabla C(\mathbf{x}) \right|$
			\STATE $\mathbf{p} \gets \mathbf{p} - 2 \left(\mathbf{p} \cdot \hat{ \mathbf{n} } \right) \hat{ \mathbf{n} }$
		\ENDIF
	\STATE

\ENDFOR
\STATE
\STATE \COMMENT{Full spatial step}
\STATE $ \mathbf{x} \gets \mathbf{x} + \epsilon \mathbf{p}$
\STATE
\STATE \COMMENT{Last momentum half step}
\STATE $ \mathbf{p} \gets \mathbf{p} - \frac{1}{2} \epsilon \nabla E \left( \mathbf{x} \right)$

\end{algorithmic}
\end{algorithm}

\subsection{Application to Nested Sampling}

Constrained Hamiltonian Monte Carlo (CHMC) naturally complements nested sampling by taking

\begin{align*}
p \left( \mathbf{x} \right) &\rightarrow \pi \left( \alpha \right) \\
C \left( \mathbf{x} \right) &\rightarrow \mathcal{L} \left( \alpha \right) - L.
\end{align*}

\noindent The CHMC samples are then exactly the samples from the constrained prior necessary for the generation of the nested samples.  A careful extension of the constraint also allows for the addition of a limited support constraint, making efficient nested sampling with, for example, gamma and beta priors immediately realizable. 

Initially, the $n$ independent samples are generated from $n$ Markov chains seeded at random across the full support of $\pi \left( \alpha \right)$.  After each iteration of the algorithm, the Markov chain generating the nested sample is discarded and a new chain is seeded with one of the remaining chains.  Note that this new seed is guaranteed to satisfy the likelihood constraint and the resultant CHMC will have no problems bouncing around the constrained distribution to produce the new sample needed for the following iteration.

A suite of C++ classes implementing nested sampling with CHMC has been developed and is available for general use.\footnote {\url{http://web.mit.edu/~betan/www/code.html}}

\subsection{Conclusions}

Constrained Hamiltonian Monte Carlo is a natural addition to nested sampling, the combined implementation allowing efficient and powerful inference for any problem with a smooth likelihood.  

\subsection{Acknowledgements}
I thank Tim Barnes, Chris Jones, John Rutherford, Joe Seele, and Leo Stein for insightful discussion and comments.

\end{document}